\documentclass[aip,jcp,reprint]{revtex4-1}

\usepackage{epsfig,graphicx,latexsym}
\usepackage{amssymb,amsmath}
\usepackage{color}
\usepackage{hyperref}
\usepackage{mathtools}

\newcommand{\ve}[1][K]{\mathbf{#1}}
\DeclareMathOperator\arccosh{arccosh}

\DeclareMathOperator\cotan{cotan}
\begin{document}
\author{M. Mangeat}
\author{T. Gu\'erin}
\author{D. S. Dean}
\affiliation{Laboratoire Ondes et
Mati\`ere d'Aquitaine (LOMA), CNRS, UMR 5798 / Universit\'e de  Bordeaux, F-33400 Talence, France}

\title{Dispersion in two-dimensional periodic channels with discontinuous profiles}

\begin{abstract}
The effective diffusivity of Brownian tracer particles confined in periodic micro-channels is smaller than the microscopic diffusivity due to entropic trapping. Here, we study diffusion in two-dimensional periodic channels whose cross-section presents singular points, such as  abrupt changes of radius or the presence of   thin walls,  with  openings, delimiting periodic compartments composing the channel. Dispersion in such systems is analyzed 
using the Fick-Jacobs' approximation. This approximation assumes a much faster equilibration in the lateral than in the axial direction, along which the dispersion is measured. If the characteristic width $a$ of the channel is much smaller than the period $L$ of the channel, {\em i.e.} $\varepsilon = a/L$ is small, this assumption is clearly valid for Brownian particles.
For discontinuous channels, the Fick-Jacobs' approximation is  only valid at the lowest order in $\varepsilon$  and provides a rough, though on occasions rather accurate, estimate of the effective diffusivity. Here we provide formulas for the effective diffusivity in discontinuous channels that are asymptotically exact at the next-to-leading order in $\varepsilon$. Each discontinuity  leads to a reduction of the effective diffusivity. We show that our theory is consistent with the picture of  effective {\em trapping rates} associated with each discontinuity, for which our theory provides explicit and asymptotically exact formulas. Our analytical predictions are confirmed by numerical analysis. Our results provide a precise quantification of the  kinetic entropic barriers associated with profile singularities. 
\end{abstract}
 
\maketitle

\section*{Introduction}

Characterizing the dispersion  of random walkers in complex heterogeneous media is an important issue that appears in contexts as various as mixing \cite{leBorgne2013stretching,dentz2011mixing,barros2012flow}, sorting \cite{bernate2012stochastic}, contaminant spreading \cite{brusseau1994transport,deanrev} and diffusion controlled reactions \cite{Condamin2007}. In particular, the dispersion of Brownian particles in channels is a paradigm for diffusion in confined and crowded environments such as biological cells, zeolites, porous media, ion channels and microfluidic devices \cite{burada2009diffusion,malgaretti2013entropic,bressloff2013stochastic,holcman2013control}. 
The relation between confining geometry and effective diffusivity has been extensively investigated in the physics and chemistry literature over the last decade \cite{yang2017hydrodynamic,burada2008entropic,reguera2006entropic,kalinay2006corrections,malgaretti2013entropic,Malgaretti2016,yang2017hydrodynamic}. One of the most popular theoretical approaches to diffusion in channels is the so-called Fick-Jacobs' (FJ) approximation \cite{jac1967}, based on a dimensional reduction. In the case of two-dimensional channels of local radius $R(x)$, with $x$ the coordinate in the longitudinal direction, the FJ approach reduced the study of tracer dispersion to that of  an effective one-dimensional particle, with  position $x(t)$,  diffusing in an effective entropic potential $\phi(x)=-k_BT\ln R(x)$. In symmetric periodic channels, the late-time effective diffusivity $D_e$ for this one-dimensional problem can then be deduced from the Lifson-Jackson formula\cite{lifson1962self},
\begin{align}
D_e\simeq D_{\mathrm{FJ}}=\frac{D_0}{\langle R\rangle\langle R^{-1}\rangle}\label{JF0},
\end{align}
where $D_0$ is the microscopic diffusivity, and $ \langle R\rangle=\int_0^L dx R(x)/L$ denotes the uniform average over the channel period $L$.  

This basic FJ approximation is valid when the typical equilibration time in the lateral direction is much smaller than the characteristic time scale of the dynamics in the longitudinal direction. This means that the FJ expression (\ref{JF0}) can be seen as the leading order term of an expansion of $D_e$ in powers of the small parameter $\varepsilon\equiv a/L$, where $a$ is the typical lateral channel width \footnote{Note that Ref.\cite{kalinay2006corrections} uses $D_\parallel/D_\perp$  as the small parameter, where $D_\perp$ and $D_\parallel$ are, respectively, the local diffusion coefficients in the lateral and longitudinal directions. Expansions in this parameter or in powers of $\varepsilon$ are equivalent. Note also that our parameter $\varepsilon$ is proportional to $\epsilon^{1/2}$ used in Ref.\cite{dorfman2014assessing}.}.   
For non-vanishing $\varepsilon$, FJ theories can be made more accurate by introducing a position dependent local diffusion coefficient $D(x)$ in the effective one-dimensional description \cite{zwanzig1992diffusion,reguera2001kinetic,kalinay2006corrections,kalinay2005extended,kalinay2005projection,kalinay2010mapping,martens2011entropic,bradley2009diffusion,berezhkovskii2011time,dagdug2012projection,valdes2014fick}. 
At  next-to-leading order, $D(x)\simeq D_0(1-R'^2/3)$ \cite{zwanzig1992diffusion,reguera2001kinetic,kalinay2006corrections},  leading (again using the Lifson-Jackson formula\cite{lifson1962self}) to 
\begin{align}
D_e\simeq \frac{D_0}{\langle R\rangle\langle R^{-1}\rangle}\left(1-\frac{\langle R'^2/R\rangle}{3 \langle R^{-1}\rangle}+\mathcal{O}(\varepsilon^4)\right) \label{Deexp2},
\end{align}
where the prime denotes the derivative with respect to $x$. For smooth channels, it has been checked\cite{dorfman2014assessing,mangeat2017dispersion} that the above formula is exact at order $\varepsilon^2$, and it can be extended to higher orders \cite{kalinay2006corrections,dorfman2014assessing,mangeat2017dispersion}. However, in the case of channel profiles presenting a discontinuity, it is straightforward to see that the next order correction  to the dispersivity $D_e$ given in Eq.~(\ref{Deexp2}) diverges. The appearance of such a divergence usually suggests two possibilities. Firstly it could be that the basic perturbation series needs to be resummed, for instance, on resummation, divergent terms appear in a denominator rather than a numerator and thus give finite contributions. The other possibility is that the true perturbation series is not analytic in the naive expansion parameter, which  in the approaches mentioned above turns out to be $\varepsilon^2$. In our study we show that it is the latter phenomenon which is at play and that the perturbation expansion parameter is in fact $\varepsilon$ rather than $\varepsilon^2$.

To treat this problem, existing approaches assume that the effective dynamics for $x(t)$ should include local traps at the points of discontinuity, the associated trapping rates are calculated approximately via the boundary homogenization approximation\cite{berezhkovskii2009one,antipov2013effective}. Recently\cite{kalinay2010mapping}, this theory has been found to be consistent with the approaches assuming a local diffusivity $D(x)$. However, the effective dispersivity contains coefficients which  not known in explicitly. In a third class of approaches, dispersion has been estimated by using   first passage arguments\cite{marchesoni2010mobility,borromeo2010particle}, which is valid in the limit of small openings between pores, but whose link with the FJ regime is unclear\cite{mangeat2017geometry}. 

The aim of the present paper is to derive a formula for the effective dispersion in discontinuous channels that is \textit{asymptotically exact} in the slowly varying limit $\varepsilon\to0$. We consider two dimensional periodic channels which possess a finite number $n$  of discontinuities in each period. Our main result is that the  dispersion in such channels can be written as  
\begin{align}
D_e \simeq \frac{D_0}{\langle R\rangle\langle R^{-1}\rangle} \left( 1- \sum_{i=1}^n \frac{\mu_i }{ L \langle R^{-1} \rangle} +\mathcal{O}(\varepsilon^2) \right). \label{Deexpgamma}
\end{align}
Here,  the positive coefficients $\mu_i$ only depend on the geometry of the $i^\mathrm{th}$ discontinuity (see below). Furthermore the effect of each distinct discontinuity is additive and thus the result applies to a wide range of channels in a 
simple, building block, like manner. The above formula generalizes  Eq.~(\ref{Deexp2}) to the case of discontinuous profiles, and shows that the associated corrections to dispersion are of order $\varepsilon$, and are as such thus much more important than for smooth channels (where they are of order $\varepsilon^2$). Importantly, \textit{our approach does not rely on a reduction of dimensionality}: we do not need to define a local diffusion coefficient near the singular parts of the channel to obtain it, such local diffusion coefficient would clearly be ill-defined near the profile discontinuity. Our analysis is however compatible with the notion of associated \textit{trapping rates} to model the singularity, and provides a means to obtain asymptotically exact formulas for such trapping rates, which are shown to be proportional to $1/\mu$. 

\begin{figure}
\centering
\includegraphics[width=8cm,clip]{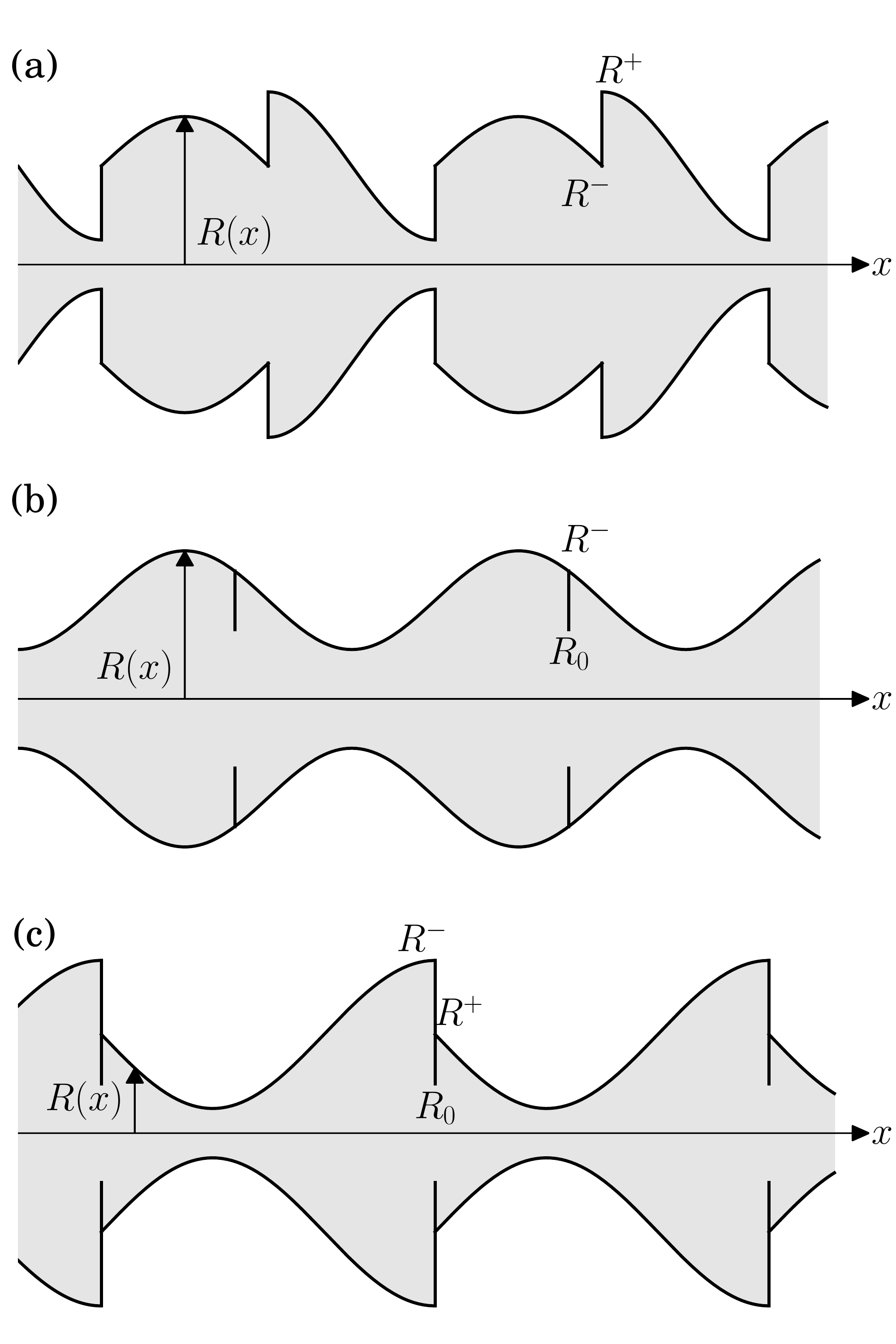}
\caption{Illustration of the three types of discontinuities of periodic channels that are considered in this paper:  (a) discontinuity of the channel radius, (b) presence of  {\em walls} separating between compartments and (c) general case composed by both type of discontinuities. \label{figTypesDiscontinuities}}
\end{figure}

\begin{figure}
\centering
\includegraphics[width=8cm,clip]{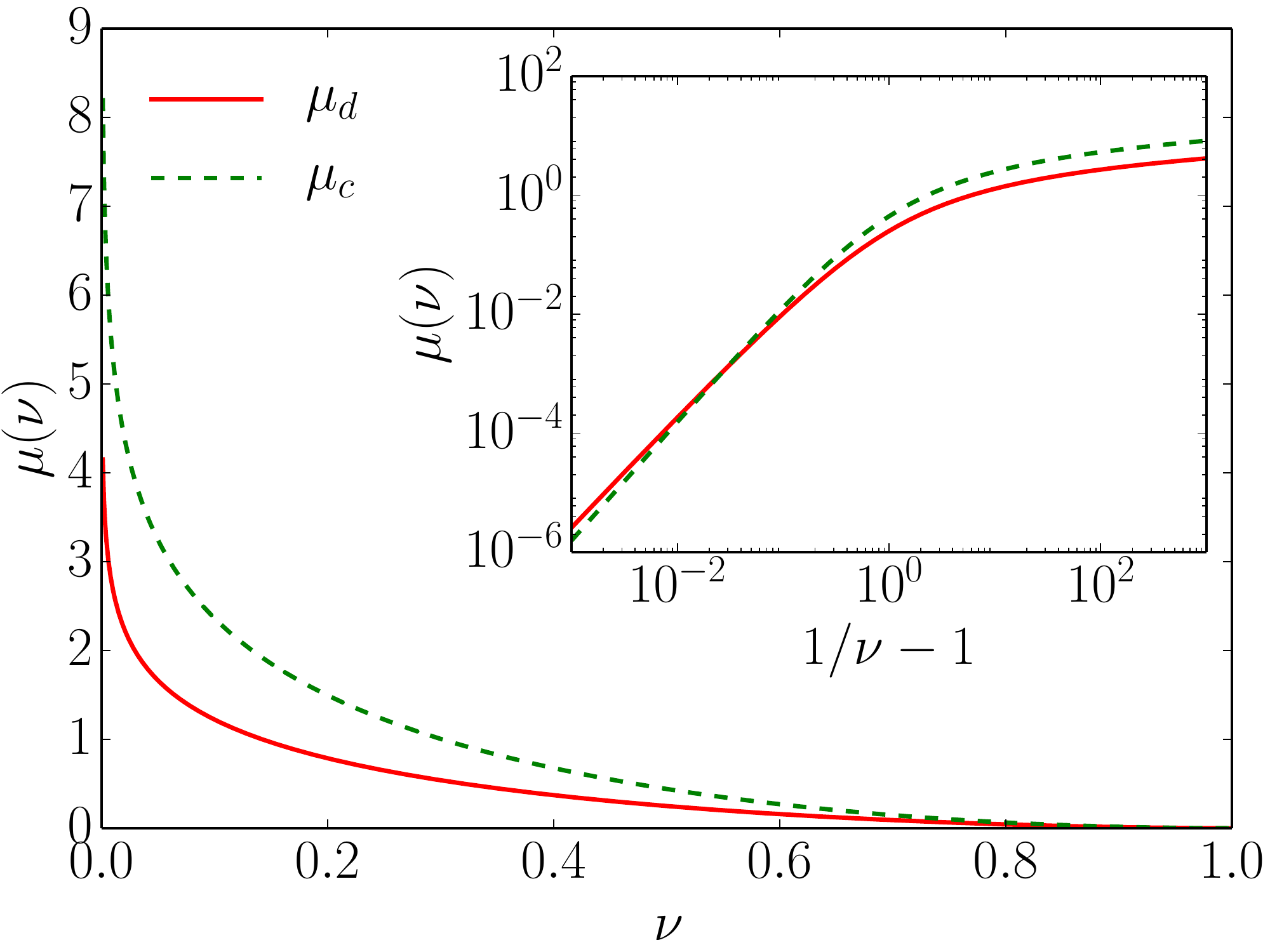}
\caption{Representation of the function $\mu(\nu) $ which quantifies the impact on dispersion of the presence of a radius discontinuity ($\mu_c$) or of a partially obstructing wall delimiting compartments ($\mu_d$). The geometric parameter is $\nu=R^-/R^+$ for discontinuous channels and $\nu=R^-/R_0$ for compartmentalized channels, where $R^-$, $R_0$ and $R^+$ are as shown in Fig.~\ref{figTypesDiscontinuities}. 
\label{figGamma}}
\end{figure}

Our formula (\ref{Deexpgamma}) shows that each discontinuity has a negative contribution to the dispersion, confirming that it effectively acts as a local trap for the Brownian particles. We have exactly calculated the coefficients $\mu$, that quantify the impact on dispersion, for two different types of basic discontinuities, shown in Fig.~\ref{figTypesDiscontinuities}. First, we have considered the case where the channel radius changes locally from a value $R^-$ to $R^+$ (see Fig.~\ref{figTypesDiscontinuities}a). In this case, $\mu$ is denoted by $\mu_d$ (where $d$ stands for discontinuous) and depends only on the parameter $\nu=R^+/R^-$:
\begin{align}
\mu_d(\nu)=\frac{1+\nu^2}{\nu \ \pi} \ln \left\vert \frac{1+\nu}{1-\nu} \right\vert   - \frac{2}{\pi} \ln \left\vert \frac{4\nu}{1-\nu^2}\right\vert.  \label{gammaexpdisc}
\end{align}
Notice that  $\mu_d(\nu) = \mu_d(\nu^{-1})$, this must be the case as we have the same diffusion constant upon flipping the direction of the channel and thus switching $R^+$ and $R^-$. 
We have also considered a second type of discontinuity, in which 
the profile contains walls that partially obstruct the channel, forming different {\em compartments} (see Fig.~\ref{figTypesDiscontinuities}b). In this case, $\mu$  is denoted by $\mu_c$ ($c$ standing for compartments) and depends on the geometric parameter $\nu=R_0/R^-$ (where $R_0$ is the radius at minimal opening and $R^-$ is the radius just next the wall), and is given by
\begin{align}
\mu_c(\nu)=-\frac{4}{\pi}\ln \left(\sin \frac{\pi \nu}{2}\right). \label{gammaexpsep}
\end{align}
Both functions $\mu_c$ and $\mu_d$ are plotted in Fig.~\ref{figGamma}. We also consider a hybrid case where the discontinuity is a combination of these type of discontinuities.

The outline of this paper is as follows. In Section \ref{SectionFormalism} we  present the general formalism used and  show that the effective diffusivity can be computed via a partial differential equation for an auxiliary function  over one channel period. In Section \ref{SectionDiscontinuous}, we consider discontinuous channels. We present a method to compute this auxiliary function with matched asymptotic expansions  and we compute  the effective diffusivity. In Section \ref{SectionSeptate}, we show how to adapt the calculation for compartmentalized channels and in Section \ref{SectionGeneral} we generalize the result to systems having hybrid forms of the   discontinuous and compartmentalized singularities. Our formulas are validated by comparison with the  numerical solutions of the relevant partial differential equations in Section \ref{SectionNumerical}. In Section \ref{SectionTrappingRates}, we determine the exact expressions for the of the trapping rates that should be  used for the boundary homogenization method, and compare them with existing approximations.

\section{General formalism: exact expression of the effective diffusivity}
\label{SectionFormalism}
We consider a symmetric two-dimensional channel of local radius $R(x)$, where $x$ denotes the longitudinal coordinate. The channel is periodic in $x$ with period $L$. We denote by  $a$ the channel width at its minimum, and we define the dimensionless profile $\zeta$ by
\begin{align}
R(x)=a\ \zeta(x/L),
\end{align}
where $\zeta(X)$ is a periodic function of $X=x/L$, with unit period, that describes the geometry of the profile.

We aim to calculate the effective diffusivity 
\begin{align}
D_e=\underset{t\rightarrow\infty}{\lim}\overline{[x(t)-x(0)]^2}/2t,
\end{align}
where $\overline{\ \cdot\ }$ represents the ensemble average over particle trajectories. Rather than reducing the problem to an effective one-dimensional dynamics for $x(t)$,  we use  the following exact expression of the effective diffusivity\cite{guerin2015,Guerin2015Kubo,mangeat2017dispersion}
\begin{equation}
\frac{D_e}{D_0} = 1- \frac{1}{\vert V \vert} \int_{\partial V} dS \ n_x \ f. \ \label{ExprDeGeneral}
\end{equation}
Here, the integral is performed on the boundary $\partial V$ of the channel (over one periodic cell), $D_0$ is the microscopic diffusivity, $dS$ represents the surface element, $n_x$ is the $x$ component of the unit normal vector $\ve[n]$ (oriented towards the exterior of the channel), $\vert V\vert=2\langle R\rangle L$ is the volume of one channel period. Furthermore, $D_e$ depends on an auxiliary function $f(x,y)$ which satisfies the Laplace equation
\begin{equation}
\nabla^2f=\partial_x^2 f + \partial_y^2 f=0 \label{Eqf},
\end{equation}
where $y$ is the transverse coordinate. In addition. $f$ is a periodic function of $x$, and  at the channel boundary it obeys the boundary condition
\begin{equation}
[\ve[n]\cdot\nabla f-n_x]_{y=\pm R(x)}=0 \label{EqBC}.
\end{equation}

The above expressions are a particular case of the formulas for the effective diffusivity for general periodic systems\cite{guerin2015,Guerin2015Kubo}, and are also consistent with the macrotransport theory of Brenner and Edwards \cite{brenner2013macrotransport}. An important dimensionless parameter of the problem is the ratio of lateral to longitudinal length  scales
\begin{align}
\varepsilon= a / L 
\end{align}
and we will study the limit of slowly varying channels, {\em i.e.} $\varepsilon \to 0$. For smooth channels, $D_e$ can be systematically expressed as an expansion in powers of $\varepsilon$. Here we focus on non-smooth channels, for which only the leading order result is exactly known  [Eq.~(\ref{JF0})].

To simplify notation, without loss of generality, we set the period length to $L=1$ and the microscopic diffusivity to $D_0=1$. In these units, $\varepsilon$ is just the typical lateral dimension of the channel, and $R=\varepsilon\zeta$. 
 
\section{Dispersion in weakly-varying discontinuous channels}
\label{SectionDiscontinuous}

Here, we first consider the case that $\zeta(x)$ presents a single discontinuity, whose origin is set at the origin $x=0$ (modulo the period). There, $\zeta(x)$ is assumed to change sharply from $\zeta^- \equiv \zeta(0^-)$  to $\zeta^+ \equiv \zeta(0^+)$, as in Fig.~\ref{figTypesDiscontinuities}a. In the slowly-varying limit $\varepsilon\to0$, characterizing $f$ is a singular perturbation problem, and it is crucial to distinguish between a region near the channel discontinuity (called the inner region), and a region far from the discontinuity (called the outer region). 

\subsection{The solution far from the discontinuity}
\label{SectionFarDisc}

We first describe the expansion of $f$ in the outer region, where it is convenient to use the rescaled variables 
\begin{align}
Y=y / \varepsilon,\hspace{1cm}X=x, 
\end{align}
so that the  ranges of $X,Y$ become independent of $\varepsilon$. We define the function $F$ such that
\begin{align}
f(x,y)= F(X,Y).
\end{align} 
It is important to note that $F$ is periodic but may present an irregular behavior (to be determined below) near the discontinuity, at $X=0$ (modulo 1). The equation satisfied by $F$ in the bulk follows from Eq.~(\ref{Eqf}),
\begin{align}
\varepsilon^2 \partial_X^2 F + \partial_{Y}^2 F=0  \label{EqF},
\end{align}
and the boundary conditions Eq. (\ref{EqBC}) become
\begin{align}
&[\varepsilon^2 \zeta'(X)\partial_XF-\partial_{Y} F]_{Y=\pm \zeta(X)} = \varepsilon^2 \zeta'(X). \label{BC_F}
\end{align}
In the limit $\varepsilon\rightarrow 0$, we look for solutions of the form
\begin{equation}
F(X,Y) = \sum_{i=0}^\infty \varepsilon^i F_i(X,Y) \label{ExpOut}.
\end{equation}
Note that here it is essential to use $\varepsilon$ as the small parameter, and not $\varepsilon^2$ which is the relevant small parameter used to study \textit{smooth} channels\cite{dorfman2014assessing}. 

Inserting this series expansion into the above equations leads to recurrence equations for the functions $F_i$. This calculation is very similar to the approach presented by Dorfman and Yariv\cite{dorfman2014assessing}, and  the details are given in Appendix \ref{AppendixFi}. At leading order, we find 
\begin{align}
F_0'(X)=1-\frac{1}{\langle \zeta^{-1}\rangle\zeta(X)}\label{F0Prime}.
\end{align}
At next order, we find that $F_1$ is discontinuous at $0$ (and hence at $1$ by periodicity)  and its derivative is given by
\begin{align}
F_1'(X) =  \frac{F_1(0^-)-F_1(0^+)}{\langle \zeta^{-1}\rangle \zeta(X)}. \label{F1Prime}
\end{align}
The unknown value of the jump $F_1(0^-)-F_1(0^+)$ will be deduced from the matching condition with the inner solution in the next section. 

\subsection{The solution near the discontinuity}
\label{SectionInner}

We now consider the function $f$ near the channel discontinuity (located at $X=0$ modulo $1$). In this region, the relevant length scale for the transverse direction is the channel width $\sim \varepsilon$. Since the change of profile is abrupt, we expect that $f$ varies with the same length scale in the longitudinal direction. This suggests that the relevant variables in the inner region are $\tilde x$ and $\tilde y$ defined by 
\begin{align}
\tilde x=x/\varepsilon= X/\varepsilon, \hspace{1cm}
\tilde y=y/\varepsilon=Y. 
\end{align}
We note that, if $\vert x\vert \ll 1$, one can simplify the domain by noting that $R(x)\simeq \varepsilon \zeta^+$ for $x>0$, and $R(x)\simeq \varepsilon \zeta^-$ for $x<0$. It is convenient to introduce the function $\phi$, defined by
\begin{align}
\phi(\tilde x,\tilde y)=f(x,y)-x. 
\end{align}
This function $\phi$ satisfies the Laplace's equation,
\begin{align}
(\partial_{\tilde x}^2+\partial_{\tilde y}^2)\phi=0,
\end{align}
and it follows from Eq.~(\ref{EqBC}) that Neumann conditions $\ve[n]\cdot \tilde\nabla\phi=0$ hold at the channel boundary. We again look for an expansion of the form
\begin{align}
\phi(\tilde x,\tilde y)=\phi_0(\tilde x,\tilde y)+  \varepsilon\phi_1(\tilde x,\tilde y)+... \label{ExpIn}
\end{align}
As a result all the  functions $\phi_i$ satisfy Laplace's equation with Neumann boundary conditions at the channel boundary, but  an additional condition is needed to determine them. This additional condition comes from the requirement that both expansions (\ref{ExpOut}) and (\ref{ExpIn}) must lead to the same value of $f$  when $\varepsilon\ll \vert x\vert \ll 1$. Thus the value of $F$ for small $X$ must be equal to $x+\phi$ when   $\tilde x\rightarrow\pm\infty$. This condition can be written explicitly as
\begin{align}
\phi+\varepsilon\tilde x \underset{\tilde x\rightarrow\pm \infty}{\simeq} F_0(0 )+\varepsilon [\tilde x F_0'(0^\pm)+F_1(0^\pm)]+...\label{954}
\end{align}
At leading order in $\varepsilon$, the above equations imply that $\phi_0\rightarrow F_0(0)$ for $\tilde x\rightarrow\pm\infty$,  the solution for $\phi_0$ is thus simply the uniform solution $\phi_0=F_0(0)$. At order $\varepsilon$, using the equations (\ref{954}) and (\ref{F0Prime}), we see that the asymptotic behavior of $\phi_1$ is
 \begin{align}
\phi_1(\tilde x\rightarrow\pm\infty,\tilde y)=F_1(0^\pm)-\frac{\tilde x}{\zeta^\pm\langle\zeta^{-1}\rangle} \label{BCP}.
\end{align}

\begin{figure}
\centering
\includegraphics[width=8cm,clip]{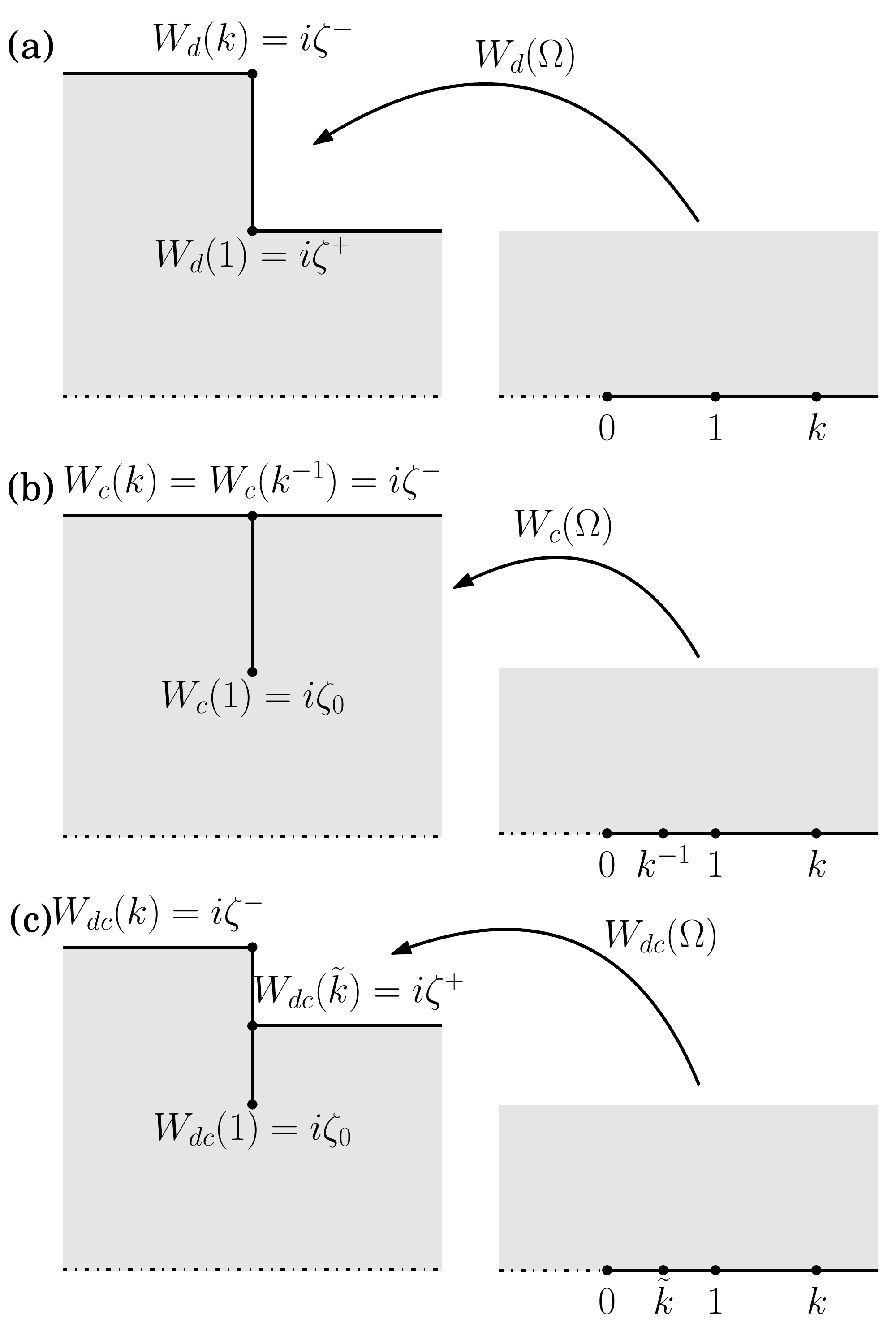}
\caption{Transformation of the boundaries close to the discontinuity for discontinuous (a), compartmentalized (b) and discontinuous-compartmentalized (c) channels after the conformal mapping $W_d(\Omega)$ given by Eq.~(\ref{SCmapdisc}), $W_c(\Omega)$ given by Eq.~(\ref{SCmapsep}) and $W_{dc}(\Omega)$ given by Eq.~(\ref{SCmapgen}) respectively.  \label{figMapping}}
\end{figure}

We also note that, by symmetry, the boundary conditions at $y=-R(x)$ can be replaced by Neumann conditions $\partial_{\tilde{y}}\phi_1=0$ at the center-line $\tilde y=0$. At this stage, we are thus left with the problem of  finding an harmonic function $\phi_1$ in a \textit{corner-shaped} domain (Fig.~\ref{figMapping}a), with Neumann conditions at the channel boundary and at the centerline, the behavior of $\phi_1$ at infinity being specified by  (\ref{BCP}). 
The solution to this problem can be obtained with a complex analysis.

We introduce the complex variable $\tilde z=\tilde x+i \tilde y$, and we consider a conformal mapping 
\begin{align}
\tilde z=W_d(\Omega),
\end{align}
such that the channel boundary and its centerline are the images of, respectively, the positive and negative real axis (Fig.~\ref{figMapping}a) in the (complex) $\Omega$-plane. 
Such a mapping can be found by using the Schwarz-Christoffel method (see appendix \ref{SCmap} for details), leading to
\begin{align}
W_d(\Omega) = \frac{\zeta^-}{\pi} \Bigg\{&\frac{1}{\sqrt k} \arccosh \left[ \frac{(k+1)\Omega-2k}{(k-1) \Omega} \right]  \nonumber\\
&-\arccosh \left[ \frac{2\Omega-(k+1)}{k-1} \right] \Bigg\}+i\zeta^-, \label{SCmapdisc}
\end{align}
where  the parameter 
\begin{align}
 k=(\zeta^-/\zeta^+)^2
\end{align}
is assumed to be larger than one (without loss of generality). Note that the image of $\Omega=1$ is $W_d(1)=i\zeta^+$, the image of $\Omega=k$ is $W_d(k)=i\zeta^-$, while the image of the negative real axis is the center-line of the channel. A similar mapping has recently been used\cite{kalinay2010mapping}, but did not lead to explicit expressions of the effective diffusivity. We check in Appendix \ref{AppendixKP} that our approach is compatible with it.

Now, since $W_d$ is a conformal mapping, the function $\phi_1$ seen as a function of $\Omega_x=\text{Re}(\Omega), \Omega_y=\text{Im}(\Omega)$ must satisfy Laplace's equation, with Neumann conditions on the boundaries which are now the positive and negative real axes. The solutions are thus of the form  
\begin{align}
\phi_1=C_1+C_2\ln \vert\Omega\vert  \label{Eq0943},
\end{align}
where $C_1$ and $C_2$ are constants. These constants are determined by making explicit the relation $\tilde x=\mathrm{Re} \ {W_d(\Omega)}$ for $\tilde x\to -\infty$ (or, equivalently,  large $\Omega$) and for $\tilde x\to+\infty$ (or, equivalently, small $\Omega$). We find  
\begin{align}
\tilde x \underset{\vert\Omega\vert \rightarrow + \infty}{\simeq}\frac{\zeta^-}{\pi} \left(\frac{1}{\sqrt k} \ln \frac{\sqrt{k}+1}{\sqrt{k}-1} - \ln \frac{4\vert\Omega\vert }{k-1}  \right),
\end{align}
and
\begin{align}
\tilde x  \underset{\vert\Omega\vert \rightarrow 0}{\simeq}\frac{\zeta^-}{\pi\sqrt{k}} \left( \ln \frac{4k}{(k-1)\vert\Omega\vert} -  \sqrt k  \ln \frac{\sqrt{k}+1}{\sqrt{k}-1}\right).
\end{align}
Inserting the value of $\ln\vert\Omega\vert$ deduced from these expressions into  (\ref{Eq0943}) and comparing with (\ref{BCP}) enables the identification of $C_2$ 
\begin{align}
C_2=&\frac{1}{\pi \langle \zeta^{-1}\rangle},
\end{align}
and of the jump of $F_1$:
\begin{align}
F_1&(0^-)-F_1(0^+)=\nonumber\\
&\frac{1}{\pi\langle\zeta^{-1} \rangle}\left(\frac{1+k}{ \sqrt k} \ln \frac{\sqrt{k}+1}{\sqrt{k}-1} - 2\ln \frac{4\sqrt{k}}{k-1}\right). \label{jumpF1disc}
\end{align} 
To summarize we have obtained an exact solution for $\phi_1$, seen as a function $\Omega$ instead of $\tilde x,\tilde y$. We shall see in the next subsection that there is no need to know  $\tilde x$ as a function of $\Omega$ to obtain the effective diffusivity.

\subsection{Expression  of the effective diffusivity for a discontinuous channel}
We now use our expressions for the auxiliary function to deduce the value of the effective diffusivity. Rewriting Eq.~(\ref{ExprDeGeneral}) leads to
\begin{align}
&D_e=1+D_\text{outer} +D_\text{inner}, \\
&D_\text{outer} = \frac{1}{\langle\zeta\rangle} \int_0^1dx\zeta'(x)f(x,R(x)),\\
&D_\text{inner}=-\frac{1}{\varepsilon\langle\zeta\rangle}\int_{\varepsilon\zeta^+}^{\varepsilon\zeta^-}dy f(0,y),
\end{align}
where we have separated the contributions coming from the  {\em inner} and the {\em outer} regions. The contribution of the {\em inner} region is
\begin{align}
 D_\text{inner}=-\frac{1}{\langle \zeta\rangle}\int_{\zeta^+}^{\zeta^{-}}d\tilde y [\phi_0 +\varepsilon \phi_1(0,\tilde y)].
\end{align}
However, we remark that for any harmonic function $\phi(x,y)$, we have the relation for any closed domain $V$:
\begin{align}
&\oint_{\partial V} dS\  n_x \phi = \int_V dV \ \nabla\phi \ve[e]_x\nonumber\\
&=-\int_V dV\  x\nabla^2\phi+ \oint_{\partial V} dS \ x \ve[n] \cdot \nabla\phi  =  \oint_{\partial V} dS \ x \ve[n] \cdot \nabla\phi.
\end{align}  
Applying this formula to $V$ large but in the boundary layer and $\phi=\phi_0+\varepsilon\phi_1$, and taking into account its boundary conditions leads to
\begin{align}
D_\text{inner}=\frac{ \zeta^+[F_0(0)+\varepsilon F_1(0^+)] - \zeta^-[F_0(0)+\varepsilon F_1(0^-)]}{\langle\zeta\rangle}.  \label{Dinner}
\end{align}
In turn, the integral for $D_\text{outer}$ is dominated by the contribution coming from the outer solution (the contributions coming from the inner-solution are of higher order since $\zeta'$ vanishes in the inner region). Hence,
\begin{align}
 D_\text{outer}= \frac{1}{\langle\zeta\rangle} \int_0^1dX\zeta'(X)[F_0(X)+\varepsilon F_1(X)].
\end{align}
Integrating by parts, we obtain
\begin{align}
& D_\text{outer}=-\frac{1}{\langle\zeta\rangle} \int_0^1dX\zeta (X)[F_0'(X)+\varepsilon F_1'(X)]\nonumber\\
&+\frac{\zeta^-}{\langle\zeta\rangle} [F_0(1)+\varepsilon F_1(1^-)]-\frac{\zeta^+}{\langle\zeta\rangle} [F_0(0)+\varepsilon F_1(0^+)]. \label{Douter}
\end{align}
Collecting the results (\ref{Dinner}), (\ref{Douter}) we see that
\begin{align}
D_e =  \frac{1}{\langle \zeta\rangle\langle\zeta^{-1}\rangle} \left\{ 1 - \varepsilon \left[ F_1(0^-)- F_1(0^+)\right]\right\}, \label{DeAndJumpF1}
\end{align}
which means that $D_e$ is simply related to the jump of the function $F_1$ at the discontinuity. 
Using Eq.~(\ref{jumpF1disc}) with $k=1/\nu^2$ finally leads to 
\begin{align}
D_e = \frac{1}{\langle \zeta\rangle\langle\zeta^{-1}\rangle} \left[ 1 -   \frac{\varepsilon \mu_d(\nu)}{\langle \zeta^{-1}\rangle}\right],
\label{eqDedisc}
\end{align}
where $\mu_d$ is given by Eq.~(\ref{gammaexpdisc}). This is the announced result in the case of channels presenting discontinuities. We notice that  $\mu_d(\nu)=\mu_d(1/\nu)$, which is a consequence of  invariance of $D_e$ under the transformation $x \rightarrow -x$.

\subsection{Presence of several discontinuities}

We now consider a channel containing several discontinuities, an example is shown in Fig.~\ref{figTypesDiscontinuities}a. In this case we can decompose the channel into several outer regions where the equations (\ref{F0Prime}) and (\ref{F1Prime}) are still verified by $F_0'$ and $F_1'$ respectively, which means that the expressions for $F_0$ and $F_1$ are identical on all outer regions up to an additive constant.

Moreover, close to the discontinuity present at $x=x_i$, we see that to the leading order  $F_0(x_i^+) = F_0(x_i^-)$, which means that $F_0$ is continuous in the entire channel. At the first order of perturbation, the jump $F_1(x_i^-) - F_1(x_i^+)$ of the function $F_1$ is given by Eq. (\ref{jumpF1disc}), which depends {\em only} on the geometry of the $i^{\rm th}$ discontinuity. These conditions on $F_0$ and $F_1$ close to all singularities of the channel impose that the expressions for $F_0$ and $F_1$ do not involve any constant depending on the outer region, just a global one. Due to the relation $\int_{\partial V} dS\ n_x = 0$, the effective diffusivity  $D_e$ is independent of this global constant in Eq.(\ref{ExprDeGeneral}). 
Furthermore, all discontinuities give an additive contribution to the diffusivity  as can be seen from  Eq.(\ref{eqDedisc}) at first order in $\varepsilon$. This leads to the expression (\ref{Deexpgamma}). Let us finally note that the above only applies in the case where the discontinuities are separated by distances ${\cal O}(1)$ and when they are separated by distances 
${\cal O}(\varepsilon)$ the analysis breaks down and the full inner solution with both discontinuities must be solved.

\section{Generalization to other types of discontinuities}

\subsection{Dispersion for weakly varying compartmentalized channels}
\label{SectionSeptate}

We now show how to adapt the results  of the previous section to consider dispersion in channels with different type of profile singularities. We consider here symmetric two-dimensional channels, which are partially obstructed by walls taken to be at the position $x=0$. We refer to these kind of channels as {\em compartmentalized} ones. At the center of these walls, we assume the presence of an opening whose (reduced) radius is $\zeta_0$. We denote $\zeta^-$ the radius just after (and before) the wall, this geometry is shown in Fig.~\ref{figTypesDiscontinuities}b.

As in the case of a discontinuous channel, we distinguish between an inner and an outer region. In the outer region, the analysis is exactly the same, and  the auxiliary function satisfies Eqs.~(\ref{F0Prime}) and (\ref{F1Prime}). In the inner region, the function $f$ has the same structure, $\phi(\tilde x,\tilde y)=f(x,y)-x$ (with the same definition of the coordinates in the boundary layer). The modification of Eq.~(\ref{BCP}) for the matching condition, which gives the value of $\phi_1$ for large arguments is given by:   
\begin{align}
\phi_1 (\tilde x \rightarrow \pm \infty, \tilde y)  = F_1(0^\pm)-\frac{\tilde x}{\zeta^- \langle \zeta^{-1} \rangle}.\label{matcheqsep}
\end{align}
The function $\phi_1$ satisfies Laplace's equation in the domain drawn in Fig.~\ref{figMapping}b, with Neumann conditions at the channel boundaries and at the centerline. We apply again the Schwarz-Christoffel method to find a conformal mapping enabling to solve for $\phi_1$. We find in Appendix \ref{SCmap} that
\begin{align}
W_c(\Omega) = & \frac{2\zeta^-}{\pi} \Bigg[ \ln\left(\frac{\sqrt{k\Omega-1} + \sqrt{\Omega/k-1}}{\sqrt{(k-1/k)\Omega}}\right) - \nonumber\\
&\ln\left(\frac{\sqrt{\Omega-k} + \sqrt{\Omega-1/k}}{\sqrt{k-1/k}}\right)\Bigg]  
+i\zeta^-\label{SCmapsep},
\end{align}
where the parameter $k$ is given by
\begin{align}
k=\cotan^2\frac{\pi \zeta_0}{4\zeta^-} ,\label{defksep}
\end{align}
and is assumed to be larger than one. Note that $W_c(1/k)= W_c(k) = i \zeta^-$, $W_c(1)= i \zeta_0$ while the image of negative real axis is the centerline of the channel and the image of the positive real axis is the channel boundary (Fig.~\ref{figMapping}b).
Following the same reasoning as before, we can express $\phi_1$ as a function of the complex variable $\Omega$, as in Eq.~(\ref{Eq0943})
\begin{align}
\phi_1 = C_1 + C_2 \ln \vert \Omega\vert.
\end{align}
We can thus deduce the jump for $F_1$ from these expressions, by inverting explicitly the mapping $\tilde x= \mathrm{Re} \ W_c(\Omega)$ for  $\tilde x \rightarrow - \infty$ (or, equivalently, $\vert \Omega \vert \rightarrow \infty$) where
\begin{align}
\tilde x   \underset{\vert \Omega \vert \rightarrow \infty}{\simeq} -\frac{\zeta^-}{\pi} \ln \frac{4k\vert \Omega\vert}{(k+1)^2},
\end{align}
and for  $\tilde x \rightarrow + \infty$ (or, equivalently, small $\vert \Omega \vert$), for which
\begin{align}
\tilde x  \underset{\vert \Omega \vert \rightarrow 0}{\simeq} -\frac{\zeta^-}{\pi} \ln \frac{(k+1)^2\vert \Omega\vert}{4k}. 
\end{align}
Comparing these expressions with Eq. (\ref{matcheqsep}), we identify the jump of the function $F_1$,
\begin{align}
F_1(0^-)-F_1(0^+) = \frac{2}{\pi\langle \zeta^{-1} \rangle}\ln \frac{(k+1)^2}{4k} \label{jumpF1sep}
\end{align}
and the value of the constant $C_2$
\begin{align}
C_2=\frac{1}{\pi\langle \zeta^{-1} \rangle}.
\end{align}
We can check that Eq.~(\ref{DeAndJumpF1}) still holds here,
\begin{align}
D_e =  \frac{1}{\langle \zeta\rangle\langle\zeta^{-1}\rangle} \left\{ 1 - \varepsilon \left[ F_1(0^-)- F_1(0^+)\right]\right\}, 
\end{align}
so that the effective diffusivity is straightforwardly deduced from the jump of the function $F_1$. Setting $\nu=\zeta_0/\zeta^-$ and using the definition (\ref{defksep}) of $k$, we finally obtain 
\begin{align}
D_e \simeq   \frac{1}{\langle \zeta\rangle\langle\zeta^{-1}\rangle} \left(1 - \frac{\varepsilon \mu_c(\nu)}{\langle \zeta^{-1} \rangle} \right),
\end{align}
which is the expression for $\mu_c$ given in Eq.~(\ref{gammaexpsep}) and is the announced result for dispersion in compartmentalized channels.

\begin{figure}
\centering
\includegraphics[width=8cm,clip]{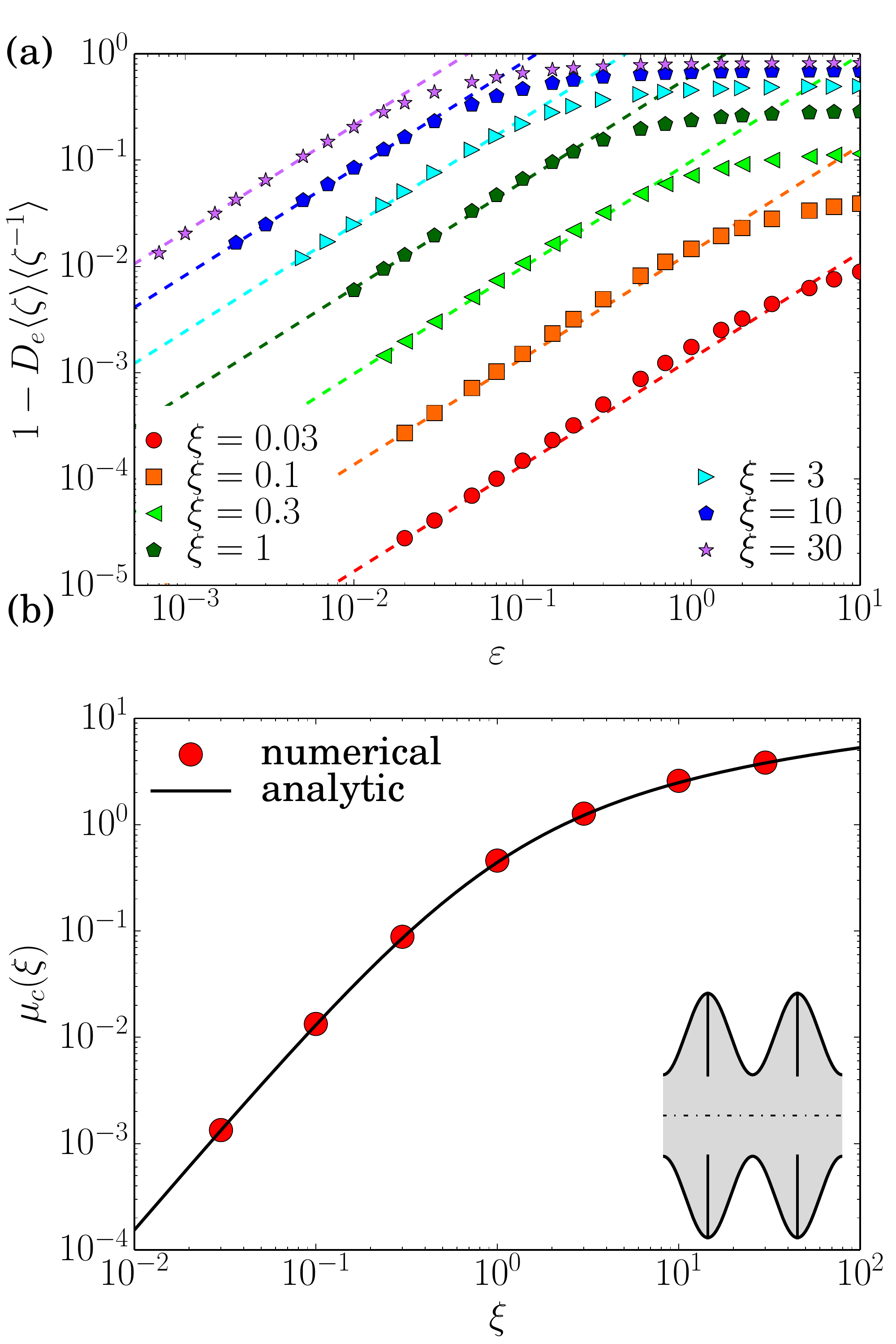}
\caption{Numerical checks of the expression (\ref{gammaexpsep}) for the effective diffusivity, in the case $\zeta(X\ne0)=1+0.5 \xi[1-\cos(2\pi X)]$, in the presence of a wall with reduced opening  radius $\zeta_0=1$ at $X=0,1,...$. Here $\nu= 1/(1+\xi)$. (a) $1-D_e/D_{\text{FJ}}$ represented for various $\xi$ [symbols: numerical solution of Eqs.~(\ref{ExprDeGeneral})-(\ref{Eqf})-(\ref{EqBC}), dashed lines: analytical prediction  (\ref{Deexpgamma})]. (b) Full line: Value of $\mu$ predicted by (\ref{gammaexpsep}), symbols: value of $\mu$ obtained from a linear fit of the data of (a) and assuming the behavior (\ref{Deexpgamma}). Inset: shape of the channel. 
The channel shape is represented in inset for $\xi=2$. \label{FigNumWall}}
\end{figure}

\subsection{The case of weakly varying discontinuous-compartmentalized channels}
\label{SectionGeneral}

We now consider the dispersion in channels with a general type of singularity mixing the two previous cases, shown in Fig.~\ref{figTypesDiscontinuities}c. Here, we assume that the channel is partially obstructed by walls (at $x=0,1,2...$) with a different radius between the negative (before the wall) and positive (after the wall) regions. We denote by $\zeta_0$ the reduced channel radius at the opening, whereas $\zeta^-$ is the radius just before the wall and  $\zeta^+$ is the radius just after the wall, as shown in the left figure of Fig.~\ref{figMapping}c. 

Following exactly the same steps as in the previous section, we obtain for this class of channels 
\begin{align}
D_e \simeq   \frac{1}{\langle \zeta\rangle\langle\zeta^{-1}\rangle} \left(1 - \frac{\varepsilon \mu_{dc}(k,\tilde k)}{ \langle \zeta^{-1} \rangle} \right),
\end{align}
Here, $\mu_{dc}$ is defined by 
\begin{align}
\label{eqgammadc}
\mu_{dc}(k,\tilde k) = \frac{1+k\tilde k}{\sqrt{k\tilde k}}  \ln \frac{\sqrt{k}+\sqrt{\tilde k}}{\sqrt{k}-\sqrt{\tilde k}} - 2 \ln \frac{4\sqrt{k\tilde k}}{k-\tilde k},
\end{align}
and the parameters $k$ and $\tilde k$ are defined by the system
\begin{align}
k\tilde k &= \left(\frac{\zeta^-}{\zeta^+}\right)^2 \\
\frac{\pi}{2} \left(\frac{\zeta_0}{\zeta^-}-1\right)&= \frac{1}{\sqrt{k\tilde k}} \arctan \sqrt{\frac{\tilde k (k-1)}{k(1-\tilde k)}} -\arctan \sqrt{\frac{k-1}{1-\tilde k}}
\end{align}
In the limit that $\nu=\zeta_0/\zeta^-\to \infty$ and fixed $\zeta_+/\zeta_0$ (\textit{i.e.} when the opening is small compared to at least one radius outside the discontinuity), we obtain the following behavior
\begin{align}
\mu_{dc}  \simeq
\begin{cases}
-\frac{4}{\pi} \ln \frac{\pi\zeta_0}{2\sqrt{\zeta^+\zeta^-}}&(\text{if } \zeta^+\ne\zeta_0)\\
\frac{2}{\pi}\left(1-\ln \frac{4\zeta^+}{\zeta^-}\right)& (\text{if } \zeta^+=\zeta_0) \label{asympteq}.
\end{cases}
\end{align}

\begin{figure}
\centering
\includegraphics[width=8cm,clip]{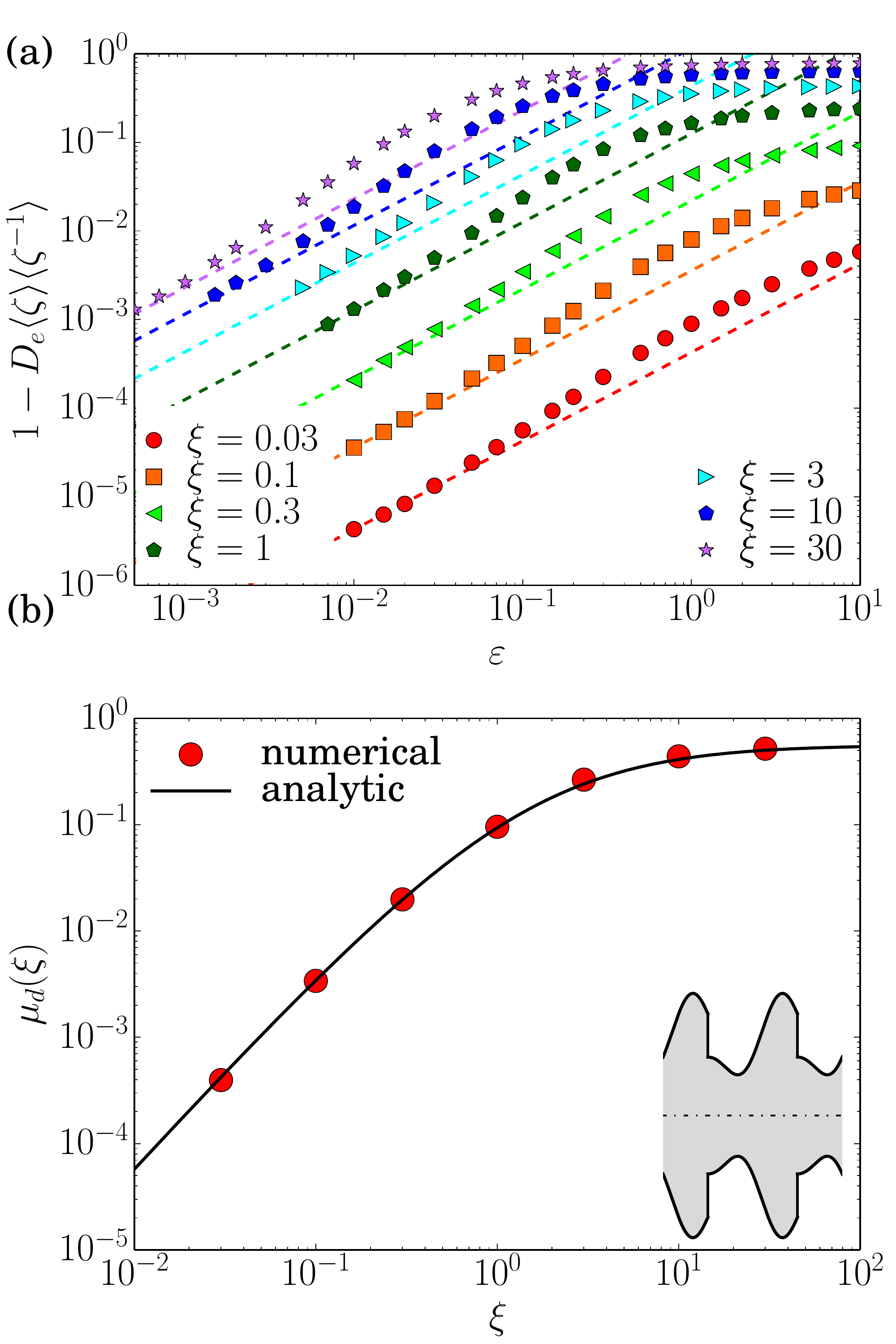}
\caption{Numerical check of the expression (\ref{gammaexpdisc}) for the effective diffusivity of discontinuous channels, whose profile is given by $\zeta(X)=1+\xi[0.39-0.41 \sin((4X+1)\pi/3)+0.20\sin((8X+1)\pi/3))]$ for $0<X<1$. Here $\nu= 1/(1+\xi)$ and $\xi$ is the amplitude of variation of the channel radius. (a) $1-D_e/D_{\text{FJ}}$ is represented for various $\xi$ [symbols: numerical solution of   Eqs.~(\ref{ExprDeGeneral})-(\ref{Eqf})-(\ref{EqBC}), dashed lines: analytical prediction  (\ref{Deexpgamma})]. (b) Full line: Value of $\mu$ predicted by (\ref{gammaexpdisc}), symbols: value of $\mu$ obtained from a linear fit of the data of (a) and assuming the behavior (\ref{Deexpgamma}). Inset: shape of the channel for $\xi=2$. \label{FigNumDisc}}
\end{figure}

\section{Comparison with numerical solutions and the literature}
\label{SectionNumerical}

We now validate our analytical approach by comparing with the exact numerical integration of the set of partial differential equations  (\ref{ExprDeGeneral})-(\ref{EqBC}). In figure \ref{FigNumDisc} we show results for an example of a discontinuous channel, whose shape is represented in the inset of Fig.~\ref{FigNumDisc}b. We first check in Fig.~\ref{FigNumDisc}a that the first corrections to the basic Fick-Jacobs' results are of order $\varepsilon$,  as opposed to smooth channels for which the correction is of order $\varepsilon^2$. Furthermore, Fig.~\ref{FigNumDisc}b shows that the coefficient of the $\varepsilon$-correction to $D_e$ is correctly predicted by our formula (\ref{gammaexpdisc}), thus validating our analytical approach. We perform a similar analysis for an example of channel presenting local walls defining compartments, represented in the inset of Fig.~\ref{FigNumWall}. The numerical analysis clearly demonstrates that the next-to-leading order term for the dispersion is correctly predicted by Eq.~(\ref{gammaexpsep}), validating our analysis for this class of channels as well.

Furthermore, the case of  discontinuous channels was considered in Ref. \cite{kalinay2010mapping}.  We check in Appendix \ref{AppendixKP} that our theory and that of Ref.~\cite{kalinay2010mapping} are consistent in the case $\zeta_-=2\zeta^+$ (which is the only case for which explicit expressions are given in Ref. \cite{kalinay2010mapping}). 

\begin{figure}
\centering
\includegraphics[width=8cm,clip]{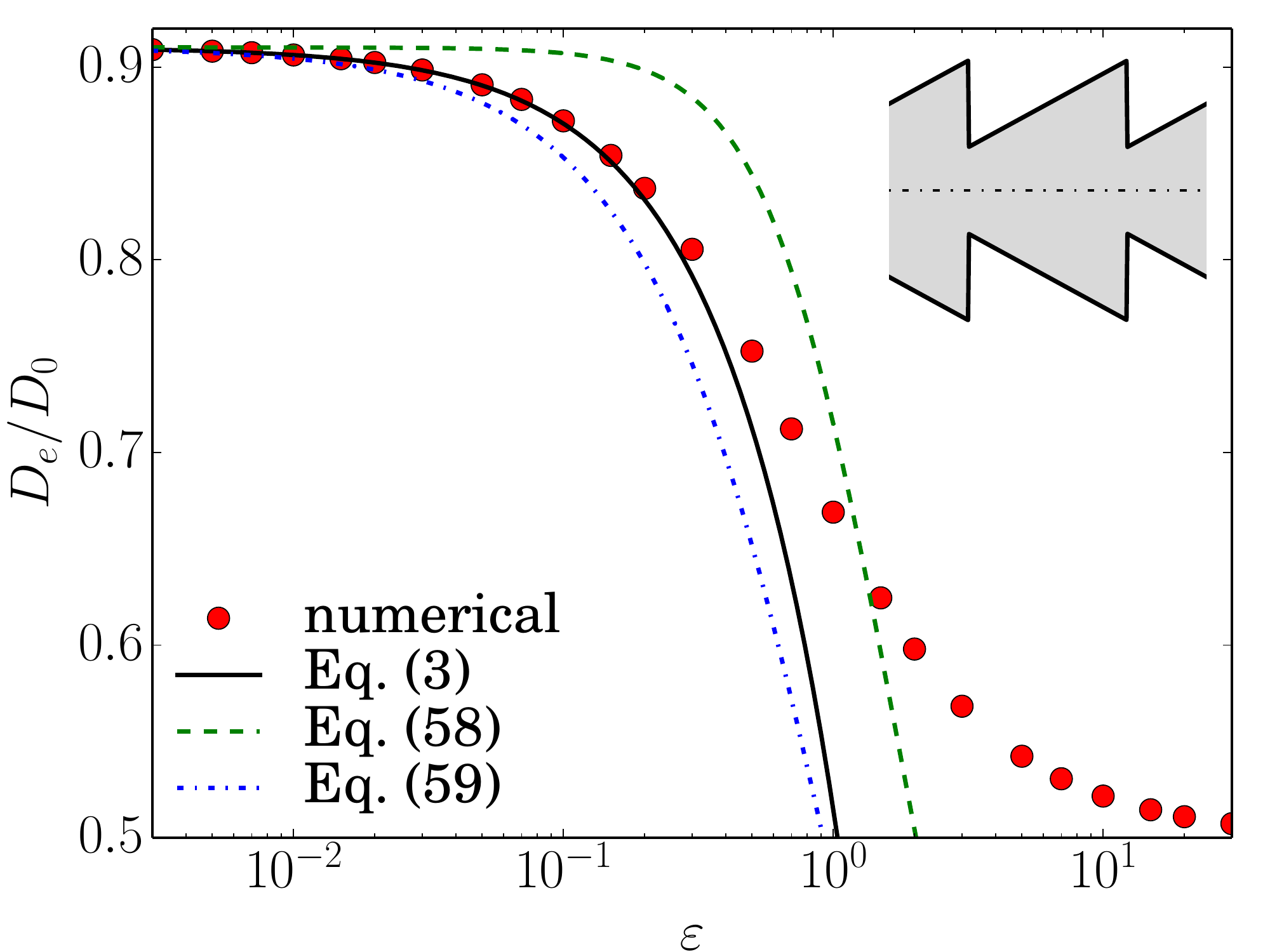}
\caption{Effective diffusivity of the ratchet channel $\zeta(x)=0.25+x$. The dots correspond to the numerical solution of the partial differential equations Eqs.~(\ref{ExprDeGeneral})-(\ref{Eqf})-(\ref{EqBC}), the straight line represents the first order correction to FJ given by  Eqs.~(\ref{Deexpgamma})-(\ref{gammaexpdisc}). We also show results obtained by using the partially resumed formula for $D(x)$  given in Ref. \cite{kalinay2006corrections} (dashed line: Eq.(\ref{EqLinearChannelChoice1}), dash-dotted line: Eq. (\ref{EqLinearChannelChoice2}), see text).  
\label{figratchet}}
\end{figure}

We can also check our analytical, asympotically-exact, result for the case of the ratchet like channel, where the profile is a periodic repetition of a linear profile $\zeta(x)=a+x$ for $x \in [0,1[$, thus presenting discontinuities at $x=0,1,2,...$. We present in Fig.~\ref{figratchet} the exact numerical integration of the set of partial differential equations (\ref{ExprDeGeneral})-(\ref{Eqf})-(\ref{EqBC}) compared to our first order correction to Fick-Jacobs given by Eq.~(\ref{Deexpgamma}). We also show  here  asymptotic results obtained by using the Kalinay and Percus\cite{kalinay2006corrections} formula for a position-dependent coefficient $D(x)=\arctan(R'(x))/R'(x)$ that is in principle exact in the linearly expanding parts of the channel. Here there are two possible  procedures to  apply the Lifson and Jackson formula\cite{lifson1962self} to determine the diffusion constant: the first where we ignore the discontinuity of the channel and find 
\begin{align}
\frac{D_e}{D_0} \simeq \frac{\arctan \varepsilon}{\langle\zeta\rangle\langle\zeta^{-1}\rangle\varepsilon},\label{EqLinearChannelChoice1}
\end{align}
which gives a correction to Fick-Jacobs result of order $\varepsilon^2$ and is thus clearly
incompatible with our exact results (see Fig.~\ref{figratchet}). Secondly the 
vertical line at the end of the channel between $y=1+a$ and $y=a$ can be replaced by
a straight line of finite slope between $(1-\delta, 1+a)$ and $(1,a)$, applying the Lifson-Jackson formula and then taking the limit $\delta\to 0$. Following this procedure leads to \cite{mangeat2017dispersion}
\begin{align}
\frac{D_e}{D_0} = \frac{1}{\langle\zeta\rangle\langle\zeta^{-1}\rangle \left(\frac{2}{\pi} \varepsilon + \frac{\varepsilon}{\arctan\varepsilon}\right)}.\label{EqLinearChannelChoice2}
\end{align}
Interestingly, this result includes a correction of order $\varepsilon$, but with a prefactor that disagrees with the exact result Eq.~(\ref{Deexpgamma}). This is not surprising since the arctangent formula for $D(x)$ is obtained by neglecting all high-order derivatives of $R(x)$ in the expansion series, whereas such terms are infinite at the discontinuity. Hence, our approach provides more precise results for this kind of channels, even if it does not include the effect of higher order  terms in the $\varepsilon$ expansion. 

\section{Effective trapping rates}

\label{SectionTrappingRates}

\begin{figure}
\centering
\includegraphics[width=8cm,clip]{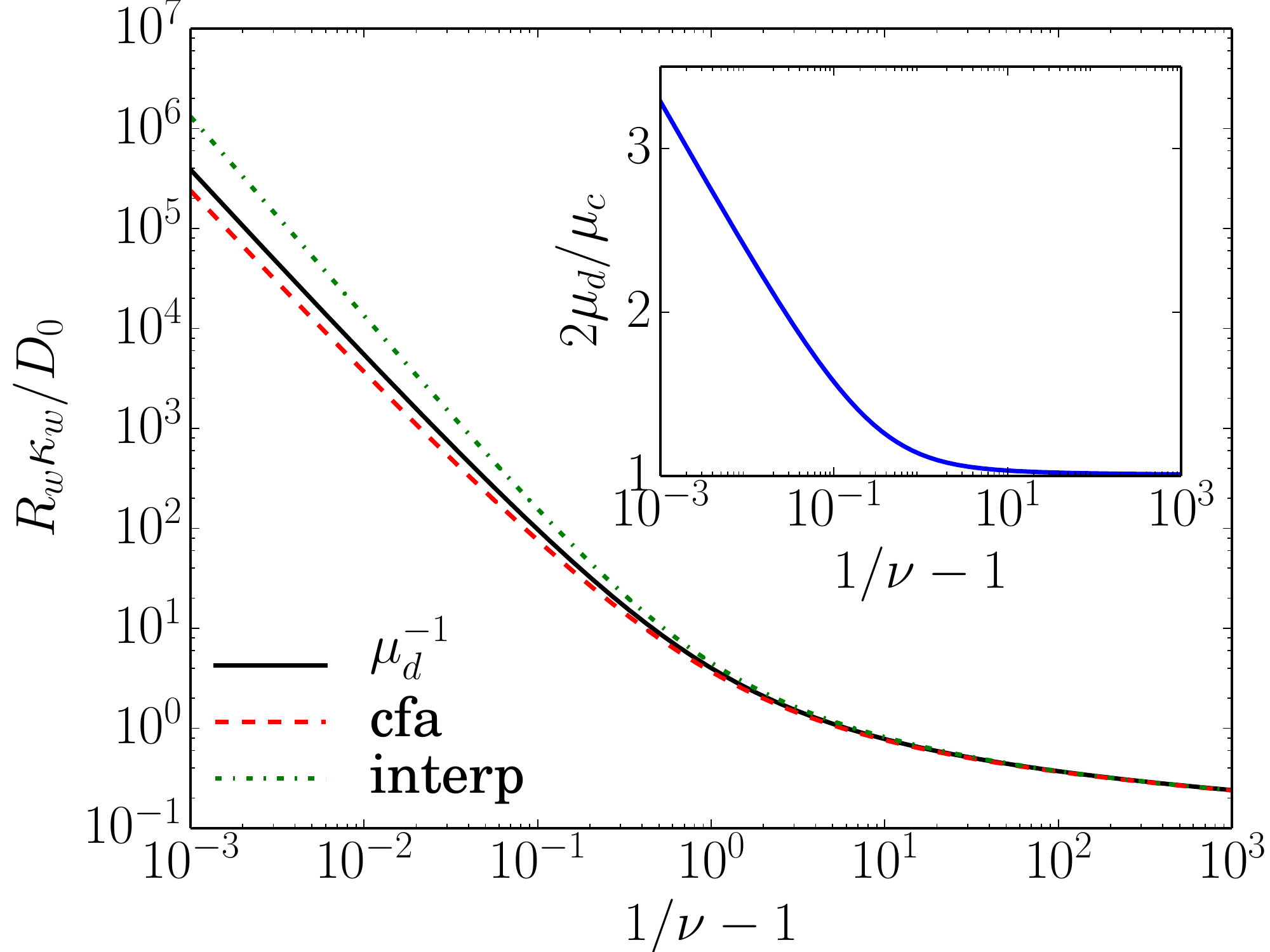}
\caption{Trapping rate $\kappa_w R_w/D_0$ as a function of $\nu=R_n/R_w$. Our theory (continuous black line) is compared to expressions for trapping rates proposed in Ref.\cite{berezhkovskii2006homogenization}, obtained by the constant flux approximation [CFA, Eq.~(\ref{trapeqcfa}), dashed red line] or interpolation [interp, Eq.~(\ref{trapeqinterp}), green dotted line].
In inset, we compare the ratio of trapping rates for the discontinuities and for compartmentalized channels, which is found to differ from unity. 
\label{figtrapping}}
\end{figure}

A widely used approach to deal with discontinuous channels is the use of the boundary homogenization method\cite{berezhkovskii2009one,makhnovskii2010diffusion,antipov2014diffusion,makhnovskii2009time,antipov2013effective}. In this class of approaches, one assumes that one can define a one dimensional stochastic dynamics for $x(t)$, with  associated probability density function $p(x,t)$ that satisfies a diffusion equation in the smooth part of the channel. The presence of discontinuities is taken into account by introducing trapping rates $\kappa^\pm$ in the flux continuity equation
\begin{align}
-D_0\partial_x p\vert_{x=0^+}=-D_0\partial_x p\vert_{x=0^-}=\kappa^-p_{x=0^-}-\kappa^+p_{x=0^+}.
\end{align}
Roughly speaking, $\kappa^-$ quantifies the likelihood, for a particle on one side of the discontinuity, to cross it (and thus be re-injected on the other side of it). The ratio of trapping rates can be deduced from detailed balance (here in the two-dimensional case)
\begin{align}
\frac{\kappa^+}{\kappa^-}=\frac{\zeta^-}{\zeta^+} \label{DetailedBalanceTrappingRates}.
\end{align}

Here we show that our approach is compatible with the concept of trapping rates, and that it provides a means to determine them exactly in the limit $\varepsilon\to 0$.  Consider first the case of a channel formed by wide ($w$) and narrow ($n$) portions of constant radius $R_w,R_n$ and length $l_w,l_n$, with $L=l_n+l_w$. One introduces two kinds of trapping rates: $\kappa_w$ quantifying the transitions from the wide to the narrow portions, and conversely $\kappa_n$ that quantifies the transitions from the narrow to the wide portions. The effective diffusivity in such channels reads (see Eq.~(31) of Ref.~\cite{makhnovskii2010diffusion})
\begin{align}
D_e = \frac{D_0L^2}{  l_n^2 + l_w^2 +l_n l_w \left(\frac{\kappa_w}{\kappa_n}+ \frac{\kappa_n}{\kappa_w}\right)  +2D_0\left( \frac{l_n}{\kappa_n}+ \frac{l_w}{\kappa_w}\right)}.
\end{align}
Using the detailed balance condition (\ref{DetailedBalanceTrappingRates}), we find  that the above formula simplifies to
\begin{align}
D_e = \frac{D_0}{\langle R\rangle\langle R^{-1}\rangle  +\frac{2D_0}{R_w\kappa_w L}\langle R\rangle  } \label{exacttraprateeq},
\end{align}
and in the weakly varying limit we obtain
\begin{align}
D_e\simeq \frac{D_0}{\langle R\rangle\langle R^{-1}\rangle}\left(1-\frac{2D_0}{R_w\kappa_w\langle R^{-1}\rangle L} +\mathcal{O}(\varepsilon^2)\right).
\end{align}
For the same channel, our approach leads to
\begin{align}
D_e \simeq \frac{D_0}{\langle R\rangle\langle R^{-1}\rangle} \left( 1- 2\frac{\mu_d}{L \langle R^{-1} \rangle} +\mathcal{O}(\varepsilon^2) \right).  
\end{align}
where the factor $2$ comes from the fact that they are two discontinuities per channel period. Comparing the above formulas gives
\begin{align}
\frac{R_w\kappa_w}{D_0} = \frac{R_n\kappa_n}{D_0}  =\frac{1}{\mu_d}. \label{kappaWExact}
\end{align}
The above formula suggests that asymptotically exact results  for the trapping rates are obtained from our analysis. 

In the boundary homogenization method, trapping rates are usually determined by considering the flux of particles on a surface presenting sticky patches. Although in most cases this method is applied to the three dimensional case, it is interesting to test its validity in the present two-dimensional situation. The corresponding problem is that of particles diffusing to a surface presenting straight strips. Two different approximate formulas\cite{berezhkovskii2006homogenization} were proposed for the trapping rate, the first one in the  {\em constant flux approximation} (CFA) leads to  
\begin{align}
\frac{R_w \kappa_w^{\text{(cfa)}} }{D_0} \simeq \frac{1}{2\nu}\frac{\pi^3\nu^3}{\sum_{n=1}^{\infty} (1/n^3) \sin^2(\pi n \nu)},\label{trapeqcfa}
\end{align}
where $\nu=R_n/R_w$. In Ref. \cite{berezhkovskii2006homogenization}, another (interpolation) formula is proposed
\begin{align}
\frac{R_w \kappa_w^{\text{(interp)}} }{D_0} \simeq \frac{\pi}{2(1-\nu)^2 \ln(2.6+0.7/\nu)}.\label{trapeqinterp}
\end{align}
It is interesting to compare these approximate values of the trapping rates with our exact calculation. We see on Fig.~\ref{figtrapping} that the three formulas give similar values for $\kappa_w$. For $\nu\to0$, all expressions of $\kappa_w$ have the same dominant behavior $1/\ln\nu$, given by Eq.~(\ref{asympteq}) for our exact value. They however differ for finite values of $\nu$, which is quantified by our approach. 

Next, in the case of channels made of a periodic arrangement of  compartments of constant radius $R_w$ and length $L$, separated by infinitely thin walls, with openings of radius $R_n$, we call $\kappa_c$ the trapping rate (which is usually called the permeability $P$), and the effective diffusivity reads\cite{makhnovskii2009time}
\begin{align}
D_e=\frac{D_0}{1+D_0/(\kappa_c L)}\simeq D_0\left(1-\frac{D_0}{\kappa_c L}\right),
\end{align}
which leads to
\begin{align}
\frac{\kappa_c R_w}{D_0}=\frac{1}{\mu_c}. \label{PExact}
\end{align}
In the literature\cite{makhnovskii2009time} it is suggested that $\kappa_c\simeq\kappa_w/2$, since  a particle that is exactly between the two compartments can switch with equal probability on each side. However, our theory clearly shows that such an argument is only an approximation: on the inset of Fig.~\ref{figtrapping}, we see that the exact ratio of $2\kappa_c/\kappa_d$, which in our theory is given by $2\mu_d/\mu_c$, is clearly different from unity. 

  


\section{Conclusion}
Let us now summarize our findings. We have calculated the effective diffusivity of non-interacting tracer particles diffusing in symmetric channels of non-uniform radius presenting singularities. In such channels, the usual Fick-Jacobs' (FJ) approach is valid at lowest order only and provides only a rough approximation of the diffusion coefficient. This is in  contrast to smooth channels, where the FJ theory can be systematically improved by taking into account higher order terms of the parameter $\varepsilon$, which measures the ratio of equilibration time in the lateral and axial directions. Here, we have identified the next-to-leading order term for the Fick-Jacobs' approach in two-dimensional discontinuous channels. We found that each discontinuity gives rise to an additive  negative correction  to the diffusion constant. This is compatible with modeling of discontinuities in terms of localized {\em trapping rates}. Our theory enables us to identify exact expressions of these trapping rates (by requiring that their use leads to asymptotically to the exact expressions of the diffusivity obtained here). The approach here provides explicit expressions for these trapping rates in terms of the  geometrical parameters of the discontinuity. Here we have considered two types of discontinuities: (i) the case of an abrupt change of radius, and (ii) the presence of thin walls with small opening that separate the channel into compartments. Our formalism could however be used to explore dispersion properties for other singularities, and can also be extended to the case of three dimensional channels. Our results help in precisely quantifying the concepts of kinetic entropic barriers associated with profile singularities. 

\appendix
\section{Calculation of the functions $F_i$}
\label{AppendixFi}
Here we describe how to calculate the functions $F_0,F_1,...$ appearing in the expansion  (\ref{ExpOut}). At order $\varepsilon^0$ and $ \varepsilon^1$, Eq.~(\ref{EqF}) becomes
\begin{align}
& \partial_Y^2 F_0=\partial_Y^2 F_1=0  
\end{align}
in the bulk, and the boundary conditions read
\begin{align}
&  \partial_{Y} F_0\vert_{Y=\zeta(X)}=\partial_{Y} F_1\vert_{Y=\zeta(X)}=0,  \\
&  \partial_{Y} F_0\vert_{Y=0}=\partial_{Y} F_1\vert_{Y=0}=0.
\end{align}
We thus deduce that the functions $F_0$ and $F_1$ do not depend on $Y$, and we denote them by $F_0(X)$ and $F_1(X)$. Examining the ${\cal O}(\varepsilon^2)$ terms in  (\ref{EqF}) yields 
\begin{eqnarray}
\partial_{Y}^2 F_2(X,Y) + F_0''(X)  =0. \label{eqo2}
\end{eqnarray}
Integrating this equation with respect to $Y$, and using $\partial_YF_2\vert_{Y=0}=0$ yields
\begin{align}
\partial_Y F_2(X,Y)=-F_0''(X)Y.  \label{95R42}
\end{align}
Now, expanding Eq.~(\ref{BC_F}) at order $\varepsilon^2$ enables us to identify the boundary condition for $F_2$ as
\begin{align}
\partial_{Y} F_2\vert_{Y=\zeta(X)} = \zeta'(X) [F_0'(X)-1],
\end{align}
which can be inserted into Eq.~(\ref{95R42}), yielding 
\begin{align}
	\zeta'(X)[F_0'(X)-1]=-F_0''(X)\zeta(X).
\end{align}
The solutions to this equation are of the form
\begin{align}
F_0'(X)=1-\frac{\lambda_0}{\zeta(X)},
\end{align}
where $\lambda_0$ is, so far, an undetermined constant. 
We can proceed further by anticipating here that $F_0$ is a continuous function at $X=0$ (modulo 1). Such property can be justified by considering the matching condition with the solution in the inner region (see the next section), and it is also justified since we do not expect that the discontinuity of the profile modifies the leading order term of the FJ approximation. With this assumption, the periodicity implies that $\lambda_0=\langle \zeta^{-1}\rangle^{-1} $ and thus
\begin{align}
F_0'(X)=1-\frac{1}{\langle \zeta^{-1}\rangle\zeta(X)}.
\end{align}
which is Eq.~(\ref{F0Prime}). 

Now, expanding at  order $\varepsilon^3$ the equations for $F$ yields
\begin{eqnarray}
& \partial_{Y}^2 F_3 + F_1''(X) =0,& \label{eqo3}\\
&\partial_{Y} F_3\vert_{Y=\zeta(X)} = \zeta'(X) F_1'(X). \label{eqo3BC}&
\end{eqnarray}
Integrating Eq.~(\ref{eqo3}) and using $\partial_Y F_3\vert_{Y=0}=0$ yields
$\partial_YF_3(X,Y)=-F_1''(X)Y$, comparing to Eq.~(\ref{eqo3BC}) we obtain 
\begin{align}
\zeta'(X) F_1'(X)= -F_1''(X) \zeta(X).
\end{align}
The solutions of this equation are of the form
\begin{align}
F_1'(X) =  \frac{\lambda_1}{\zeta(X)}.
\end{align}
where $\lambda_1$ is a constant. Note that $\lambda_1$ is related to the difference of the values on each side of the profile discontinuity by
\begin{align}
\lambda_1=\frac{F_1(1^-)-F_1(0^+)}{\langle \zeta^{-1}\rangle}=\frac{F_1(0^-)-F_1(0^+)}{\langle \zeta^{-1}\rangle},
\end{align}
where we have used the periodicity of $F$ in the second equality. $F_1'$ is thus given by
\begin{align}
F_1'(X) =  \frac{F_1(0^-)-F_1(0^+)}{\langle \zeta^{-1}\rangle \zeta(X)}.  
\end{align}
which is exactly Eq.~(\ref{F1Prime}).

\section{Details on conformal maps}
\label{SCmap}

According to the rules of the Schwarz Christoffel mapping\cite{mathews2012complex}, the complex derivative of the mapping $\tilde x=W_d(\Omega)$ in the case of a discontinuous channel (Fig.~\ref{figMapping}a) is of the form
\begin{equation}
W_d'(\Omega) = K_0 \frac{\sqrt{\Omega-1}}{\Omega\sqrt{\Omega-k}},
\end{equation}
where $K_0$ and $k$ are constants to be determined below. Integrating the above expression yields
\begin{align}
W_d(\Omega) =& K_0\Bigg\{\arccosh \left[ \frac{2\Omega-(k+1)}{k-1} \right] \nonumber\\
&- \frac{1}{\sqrt k} \arccosh \left[ \frac{(k+1)\Omega-2k}{(k-1) \Omega} \right]\Bigg\}+K_1.
\end{align}
The conditions that $W_d(1)=i\zeta^+$, $W_d(k)=i\zeta^-$ and $\mathrm{Im} \ W_d(0^-)=0$ then fix the values of $k,\ K_0$ and $K_1$, and we find
\begin{align}
&\sqrt{k}=\frac{\zeta^-}{\zeta^+}, K_1=i\zeta^-,\ K_0=-\frac{\zeta^-}{\pi}.
\end{align}
In the case of compartmentalized channels (Fig.~\ref{figMapping}b), we look for a mapping of the form
\begin{equation}
W_c'(\Omega) = K_0 \frac{\Omega-1}{\Omega\sqrt{(\Omega-k)(\Omega-\tilde k)}}.
\end{equation}
Integrating leads to
\begin{align}
W_c(\Omega) = &-2 K_0  \Bigg[\frac{1}{\sqrt{k\tilde k}}\ln\left(\frac{\sqrt{k(\Omega-\tilde k)} + \sqrt{\tilde k(\Omega-k)}}{\sqrt{(k-\tilde k)\Omega}}\right)  \nonumber\\
&- \ln\left(\sqrt{\Omega-k} + \sqrt{\Omega-\tilde k}\right)\Bigg] +K_1. \label{SCgenstart}
\end{align}
The conditions $W_c(1)=i\zeta_0$, $W_c(k)=W_c(\tilde k)=i\zeta^-$ and $\mathrm{Im} \ W_c(0^-)=0$ then determine the values of $k,\ \tilde k,\ K_0$ and $K_1$; we find
\begin{align}
&k=\frac{1}{\tilde k} = \cotan^2\frac{\pi \zeta_0}{4\zeta^-},\ K_0=-\frac{\zeta^-}{\pi},\\
&K_1=i\zeta^- + \frac{\zeta^-}{\pi} \ln(k-1/k) .
\end{align}

\section{Calculations for weakly varying discontinuous-compartmentalized channels}
\label{SectionGeneralAppendix}


Here we describe the calculations leading to the result
(\ref{eqgammadc}), for channels partially obstructed by walls at a given position $x=0$ and with a discontinuity of the radius between the negative (before the wall) and positive (after the wall) regions. The notations are those of Fig.~\ref{figMapping}c. As in the case of a discontinuous channel, we distinguish between an inner and an outer region. In the outer region, the analysis is exactly the same, and  the auxiliary function satisfies Eqs.~(\ref{F0Prime}) and (\ref{F1Prime}). In the inner region, the function $f$ has the same structure, $\phi(\tilde x,\tilde y)=f(x,y)-x$ (with the same definition of the coordinates in the boundary layer). The matching condition given by Eq.~(\ref{BCP}) is still satisfied.

The function $\phi_1$ satisfies Laplace's equation in the domain drawn in Fig.~\ref{figMapping}c, with Neumann conditions at the channel boundaries and at the centerline. We apply again the Schwarz-Christoffel method to find a conformal mapping enabling to solve for $\phi_1$. We find from the expression (\ref{SCgenstart}) that
\begin{align}
W_{dc}(\Omega) = & \frac{2\zeta^-}{\pi}  \Bigg[\frac{1}{\sqrt{k\tilde k}}\ln\left(\frac{\sqrt{k(\Omega-\tilde k)} + \sqrt{\tilde k(\Omega-k)}}{\sqrt{(k-\tilde k)\Omega}}\right)  \nonumber\\
&- \ln\left(\frac{\sqrt{\Omega-k} + \sqrt{\Omega-\tilde k}}{\sqrt{k-\tilde k}}\right)\Bigg] +i\zeta^-\label{SCmapgen},
\end{align}
Here, the parameters  $k$ and $\tilde k$ are chosen such that $W_{dc}(\tilde k) = i \zeta^+$, $W_{dc}(k) = i \zeta^-$, $W_{dc}(1)= i \zeta_0$ while the image of negative real axis is the centerline of the channel and the image of the positive real axis is the channel boundary (Fig.~\ref{figMapping}c). This leads to the system
\begin{align}
k\tilde k &= \left(\frac{\zeta^-}{\zeta^+}\right)^2, \\
\frac{\pi}{2} \left(\frac{\zeta_0}{\zeta^-}-1\right)&= \frac{1}{\sqrt{k\tilde k}} \arctan \sqrt{\frac{\tilde k (k-1)}{k(1-\tilde k)}} -\arctan \sqrt{\frac{k-1}{1-\tilde k}}.
\end{align}
Following the same reasoning as before, we can express $\phi_1$ as a function of the complex variable $\Omega$, following the Eq.~(\ref{Eq0943}). We can thus deduce the jump for $F_1$ from these expressions, by inverting explicitly the mapping $\tilde x= \mathrm{Re} \ W_{dc}(\Omega)$ for  $\tilde x \rightarrow - \infty$ (or, equivalently, $\vert \Omega \vert \rightarrow \infty$) where
\begin{align}
\tilde x   \underset{\vert \Omega \vert \rightarrow \infty}{\simeq} \frac{\zeta^-}{\pi} \left[ \frac{1}{\sqrt{k\tilde k}}  \ln \frac{\sqrt{k}+\sqrt{\tilde k}}{\sqrt{k}-\sqrt{\tilde k}} - \ln \frac{4\vert \Omega\vert}{k-\tilde k}\right],
\end{align}
and for  $\tilde x \rightarrow + \infty$ (or, equivalently, small $\vert \Omega \vert$), for which
\begin{align}
\tilde x  \underset{\vert \Omega \vert \rightarrow 0}{\simeq} \frac{\zeta^-}{\pi} \left[ \frac{1}{\sqrt{k\tilde k}}  \ln \frac{4k\tilde k}{(k-\tilde k)\Omega} - \ln \frac{\sqrt{k}+\sqrt{\tilde k}}{\sqrt{k}-\sqrt{\tilde k}}\right]. 
\end{align}
Comparing these expressions with Eq.~(\ref{BCP}) we identify the jump of the function $F_1$,
\begin{align}
F_1&(0^-)-F_1(0^+) =\nonumber\\
 &\frac{1}{\pi\langle \zeta^{-1} \rangle}    \left[ \frac{1+k\tilde k}{\sqrt{k\tilde k}}  \ln \frac{\sqrt{k}+\sqrt{\tilde k}}{\sqrt{k}-\sqrt{\tilde k}} - 2 \ln \frac{4\sqrt{k\tilde k}}{k-\tilde k}\right] .   \label{jumpF1gen}
\end{align}
We can check that Eq.~(\ref{DeAndJumpF1}) still holds here, and we finally obtain Eq.(\ref{eqgammadc}).

\section{Comparison with the Kalinay Percus approach}
\label{AppendixKP}
Here we control that our approach is consistent with the results of Kalinay and Percus\cite{kalinay2010mapping}, who mapped the dynamics of $x(t)$ on a one-dimensional diffusive dynamics, whose diffusion coefficient at the vicinity of a discontinuity at $x=0$ reads
\begin{align}
\frac{D_0}{D(x)} = R(x) \frac{d}{dx} \frac{x+ C_t \Theta(x) + C_0 }{R(x)} \label{EqD1D}
\end{align}
where $C_t$ and $C_0$ depend on $\zeta^\pm$. Let us recall here the Lifson-Jackson\cite{lifson1962self} formula which provides the effective diffusivity for one-dimensional particles with diffusion coefficient $D(x)$ moving in two-dimensional channels:
\begin{align}
D_e=\frac{1}{\langle R\rangle\langle [D(x)R(x)]^{-1}\rangle }. 
\end{align}
If we insert (\ref{EqD1D}) into the above expression, we see that for a  periodic channel, made of flat portions with radii $R_w$ and $R_n$ for respectively wide and narrow regions (as in Sec.~\ref{SectionTrappingRates}), we obtain
\begin{align}
D_e = \frac{D_0}{\langle R \rangle\langle R^{-1}\rangle + 2 \left( \frac{C_t+C_0}{R_n} - \frac{C_0}{R_w} \right)\langle R \rangle}. 
\end{align}
This formula is compatible with Eq.~(\ref{exacttraprateeq}) for an inverse trapping rate equal to
\begin{align}
\frac{D_0}{R_w \kappa_w} =  \frac{C_t+C_0}{R_n} - \frac{C_0}{R_w}.
\end{align}
From the Eq.~(\ref{kappaWExact}), we can thus identify
\begin{align}
\mu_d\left(\frac{R_n}{R_w}\right) =  \left(  \frac{C_t+C_0}{R_n} - \frac{C_0}{R_w} \right).
\end{align}
The values of $C_t$ and $C_0$ are given by Kalinay and Percus\cite{kalinay2010mapping} for the radii $R_n=\pi/2$ and $R_w=\pi$, yielding $C_t=1.21640$ and $C_0=-1.64792$. This leads to the  value of the inverse of trapping rate $D_0 / (R_w \kappa_w) \simeq 0.2498 $. For $\nu=R_n/R_w=0.5$, our approach gives $D_0 / (R_w \kappa_w) =\mu_d(\nu=0.5) \simeq 0.2498 $. Our result is thus compatible with that of  Kalinay and Percus\cite{kalinay2010mapping}  for $\nu=0.5$.
 

\end{document}